\pdfoutput=1
\documentclass[10pt,conference,letterpaper]{IEEEtran}
\IEEEoverridecommandlockouts
\usepackage{amsmath,amsfonts}
\usepackage{array}
\usepackage[caption=false,font=footnotesize,labelfont=sf,textfont=sf]{subfig}
\usepackage{textcomp}
\usepackage{stfloats}
\usepackage{url}
\usepackage{graphicx}
\usepackage[table]{xcolor}
\usepackage{cite}
\usepackage{booktabs}
\usepackage{enumitem}
\usepackage{microtype}
\newcommand{\orca}{\textsc{ORCA}}
\newcommand{\desta}{\textsc{DeSTA2.5-Audio}}
\newcommand{\qwenaudio}{\textsc{Qwen2-Audio}}

\begin{document}

\title{Escaping the Procrustean Bed:\\
Groupwise Orthogonal Connectors for Audio-Language Models}

\author{
\IEEEauthorblockN{Ho-Lam Chung\textsuperscript{1,2},
Ke-Han Lu\textsuperscript{1},
Yi-Cheng Lin\textsuperscript{1},
Guan-Ting Lin\textsuperscript{1},
Yiming Chen\textsuperscript{2},
Hung-yi Lee\textsuperscript{1}}
\IEEEauthorblockA{\textsuperscript{1}Graduate Institute of Communication Engineering, National Taiwan University, Taipei, Taiwan\\
\textsuperscript{2}ASUS AICS, Taipei, Taiwan}
}

\maketitle

\begin{abstract}
Audio-language models compress a speech encoder's output through a Querying Transformer (Q-Former) connector before feeding it to a large language model.
We identify two failures in this compression.
The connector's output vectors collapse to a single direction, and different speakers produce nearly indistinguishable outputs, with paralinguistic cues such as speaker identity, gender, and prosody lost along the way.
Our method, \orca{}, reverses this collapse by splitting the queries into groups whose outputs are constrained to point in different directions.
On SAKURA multi-hop reasoning, \orca{} gains 26.4 points over an identically trained 4B baseline, reaching 75.2\% (vs.\ 49.0\% for the 8B Audio Flamingo-3).
At the connector level, the same change cuts query redundancy by 12$\times$ and raises cross-speaker variance by 75$\times$.
\end{abstract}


\section{Introduction}
\label{sec:intro}

An audio-language model (ALM) extends a large language model (LLM) to accept audio input alongside text.
The standard pipeline has three components: an audio encoder (e.g., Whisper~\cite{radford2022whisper}) that maps audio to a sequence of hidden states, a connector that compresses this sequence into a fixed number of tokens, and an LLM (e.g., Llama~\cite{llama3}, Qwen~\cite{yang2025qwen3}) that reasons over the combined audio--text prefix.
A widely used connector design is the Q-Former~\cite{li2023blip2}, in which $K$ learnable queries cross-attend to the encoder output and produce a $K$-token summary for the LLM~\cite{tang2024salmonn, ghosh-etal-2024-gama, lu2025desta, ICLR2025_36c20807, wang25m_interspeech}.
This makes the connector an information bottleneck: it determines the upper bound of audio information available to all downstream modules, and what it discards, no amount of LLM adaptation can recover.

In practice, this compression fails to preserve much of the paralinguistic information in the signal. 
Speech carries semantic content (\textit{what} is said) and paralinguistic cues such as speaker identity, gender, and prosody (\textit{how} it is said).
Recent benchmarks show that ALMs fall well below human performance on paralinguistic reasoning~\cite{yang2025sakura,wang2025cpbench,ao2024sdeval,wang2026mmsu}, and controlled experiments reveal a systematic preference for textual over acoustic cues~\cite{chen-etal-2026-audio, xiong2026deaf}. Upstream of the connector, the encoder's intermediate representations retain measurable paralinguistic information~\cite{pang2026audiollmslistenread}. Yet in~\S\ref{sec:diagnosis}, we show that the connector's output is highly redundant, with very little variation across speakers, pointing to the compression itself as a key site of loss.

The shape of this loss has a vivid analogue in the figure of Procrustes from Greek myth.
Procrustes was an innkeeper with a single iron bed, and he forced every visitor onto it: those too tall were trimmed, those too short were stretched, so all guests left the same length as the bed.
In the ALM pipeline, the bed is the language-modeling objective: it rewards only what helps the LLM predict text.
The $K$-token bottleneck makes the fit tighter still, because with only $K$ vectors to represent the full signal, there is no spare room for information the loss does not reward.
Acoustic variation that aligns with text prediction fits the bed and survives; variation that does not is cut away. 
The result is a connector whose output is shaped by one objective alone, leaving non-semantic acoustic detail nowhere to live.

We identify two measurable symptoms of this Procrustean compression. The first is query collapse: the connector's $K$ output vectors converge to near-identical directions. The second is information loss: different speakers reading the same sentence produce nearly interchangeable connector outputs. We argue that query collapse is the upstream cause. When the output vectors cluster in one direction, the connector compresses a rich signal down to a single axis of variation, and speaker differences that depend on other acoustic dimensions are lost. This causal link yields a testable prediction: a purely geometric constraint that forces the queries apart should recover speaker variation as a downstream effect, without targeting any specific acoustic attribute.

We test it with \orca{} (Orthogonal Representation for Controllable Audio), which replaces the standard Q-Former with a groupwise orthogonal connector (Fig.~\ref{fig:overview}).
The $K$ queries are partitioned into $G$ groups, each with its own learnable queries and its own learned combination of encoder layers, and a regularizer pushes the group centers to be orthogonal in the output space.
The total parameter count and the inference cost match the baseline.

Our contributions are:
\begin{enumerate}[leftmargin=*,noitemsep,topsep=2pt]
    \item We diagnose query collapse and information loss at the connector (\S\ref{sec:diagnosis}).
    \item We propose \orca{}, a groupwise orthogonal connector that targets this collapse through a purely geometric constraint, requiring no attribute-specific supervision (\S\ref{sec:method}).
    \item Empirically, this one change increases cross-speaker variance by 75$\times$ without an explicit variance objective, and the same model gains +26.4 points on SAKURA multi-hop reasoning over an identically trained 4B baseline (\S\ref{sec:experiments}).
\end{enumerate}

\section{Related Work}
\label{sec:related}

Empirical studies of speech--text alignment have examined information flow through the audio-text interface.
The alignment objective shifts speech representations toward the geometry of text, with directional convergence increasing in deeper layers~\cite{xiang2025understanding}.
Hsu et al.~\cite{hsu2026anatomy} dissect this process layer by layer and identify an information-dilution phase at intermediate layers.
The \desta{} line~\cite{lu2024desta,lu2025desta} mitigates the resulting drift through descriptive self-generated training targets, operating on the supervision signal rather than on the connector's internal structure.
Our work takes a complementary approach: we restructure the connector's output geometry rather than its training signal.

The drift described above echoes broader collapse phenomena in the representation-learning literature.
Auto-regressive training drives language-model embeddings into a narrow, anisotropic cone, an effect termed the representation degeneration problem~\cite{gao2019representation,ethayarajh2019contextual}.
A related phenomenon, dimensional collapse, has been documented in contrastive self-supervised learning~\cite{jing2024understanding}: alignment objectives concentrate variance along a few principal directions.
These findings concern encoders and embedding layers; the connector modules that bridge modalities have received less theoretical attention, though they face the same objective-driven pressure.

Orthogonality is a standard tool for enforcing feature diversity~\cite{saxe2014exact,bansal2018can}, and the orthogonal Procrustes problem~\cite{schonemann1966generalized} casts alignment itself as finding the orthogonal map that best relates two sets of vectors.
In multimodal emotion recognition, angular constraints between shared and private spaces preserve prosodic cues~\cite{che2025angle}, and vision-language work uses orthogonal projections to keep modality-specific information from being overwritten~\cite{chaudhuri2025collapse}.
Related ideas appear in disentangled representation learning, which separates factors of variation into independent subspaces~\cite{kim2018disentangling,locatello2019challenging,cai2025mate}, and in group- or slot-structured models that factor a signal into specialized parts~\cite{xu2022groupvit,locatello2020object}.
These efforts apply orthogonality \textit{between modalities}.
\orca{} differs by applying it \textit{within} the audio connector, between groups of queries that share the same modality, and by letting each group choose its own encoder depth.

\section{Diagnosing the Procrustean Bed}
\label{sec:diagnosis}

We quantify the two failures on the \desta{} 8B baseline, built from Whisper-Large-V3, a $K{=}64$ Q-Former connector, and Llama-3.1-8B-Instruct.
All diagnostic experiments use CREMA-D~\cite{cao2014crema}, a corpus of acted emotional speech designed for paralinguistic research.
It contains 7{,}442 utterances from 91 actors, 48 male and 43 female, each reading the same 12 sentences in 6 emotion classes.
This factorial structure lets us hold one factor fixed and vary another: contrasting different sentences by the same speaker isolates semantic shifts, while contrasting different speakers reading the same sentence isolates non-semantic ones.
Emotion serves as a control to verify that the findings are stable across speaking styles.

The first diagnostic measures query collapse.
For each utterance, we compute the mean pairwise cosine similarity among the connector's $K{=}64$ output vectors; a value near 1.0 means the vectors have collapsed to a single direction.
Averaged across all utterances, the baseline gives \textbf{0.923} (Table~\ref{tab:baseline_diagnostics}).
Computed separately for each emotion or each sentence, the per-group values vary by at most 0.003, indicating that the collapse is a property of the model rather than of any particular input.

The second diagnostic measures information loss.
A connector that preserves speaker variation should produce different vectors for the same text spoken by different speakers.

Because absolute cosine similarity is not comparable across models with different embedding geometries~\cite{ethayarajh2019contextual}, we define a relative discriminative margin
\begin{equation}
    \Delta S = S_{\text{same}} - S_{\text{random}},
\end{equation}
where $S_{\text{same}}$ averages over all pairs sharing the same sentence but differing in speaker, and $S_{\text{random}}$ averages over random pairs that differ in both text and speaker.
A large $\Delta S$ means same-text pairs are more similar than random ones; $\Delta S$ near zero means they are not.
The \desta{} 8B gives $\Delta S {=} 0.010$, with $S_{\text{same}} {=} 0.925$ and $S_{\text{random}} {=} 0.915$.
The high absolute values place all utterances in a narrow region of cosine-similarity space; the near-zero margin shows that same-text pairs are not separated from random ones within it.
This is consistent with query collapse: when the output vectors are nearly collinear, all variation, whether cross-speaker or cross-sentence, is expressed along a single direction.
If query collapse drives the information loss, then fixing collapse alone should reduce it, even without a loss term that targets cross-speaker variation directly.

\begin{table}[t]
\centering
\caption{\textbf{Baseline diagnostics on CREMA-D.} Query collapse: queries are nearly indistinguishable (cosine similarity 1.0 means identical, 0 means orthogonal). Information loss: same-text-different-speaker pairs are barely more similar than random pairs, so the connector cannot separate semantic from acoustic shifts.}
\label{tab:baseline_diagnostics}
\setlength{\tabcolsep}{4pt}
\begin{small}
\begin{tabular}{lc}
\toprule
\textbf{Diagnostic on \desta{} 8B baseline} & \textbf{Value} \\
\midrule
Mean off-diagonal query cosine sim. & 0.923 \\
$S_{\text{same}}$ (same text, different speakers) & 0.925 \\
$S_{\text{random}}$ (different text and speakers) & 0.915 \\
Discriminative margin $\Delta S = S_{\text{same}} - S_{\text{random}}$ & 0.010 \\
\bottomrule
\end{tabular}
\end{small}
\end{table}

\section{Groupwise Orthogonal Connector}
\label{sec:method}

We first fix notation.
A Q-Former~\cite{li2023blip2} uses $K$ learnable query vectors $\{\mathbf{q}_1,\ldots,\mathbf{q}_K\}\in\mathbb{R}^d$.
For each query, the module computes
\begin{equation}
    \mathbf{z}_k = \mathrm{CrossAttn}(\mathbf{q}_k, \mathbf{H}, \mathbf{H}),
\end{equation}

where $\mathbf{H}\in\mathbb{R}^{T\times d_h}$ stacks $T$ frames of encoder output.
The $K$ output tokens are projected into the LLM's embedding space and prepended to the text context as an audio prefix.

\begin{figure}[t]
\centering
\includegraphics[width=\linewidth]{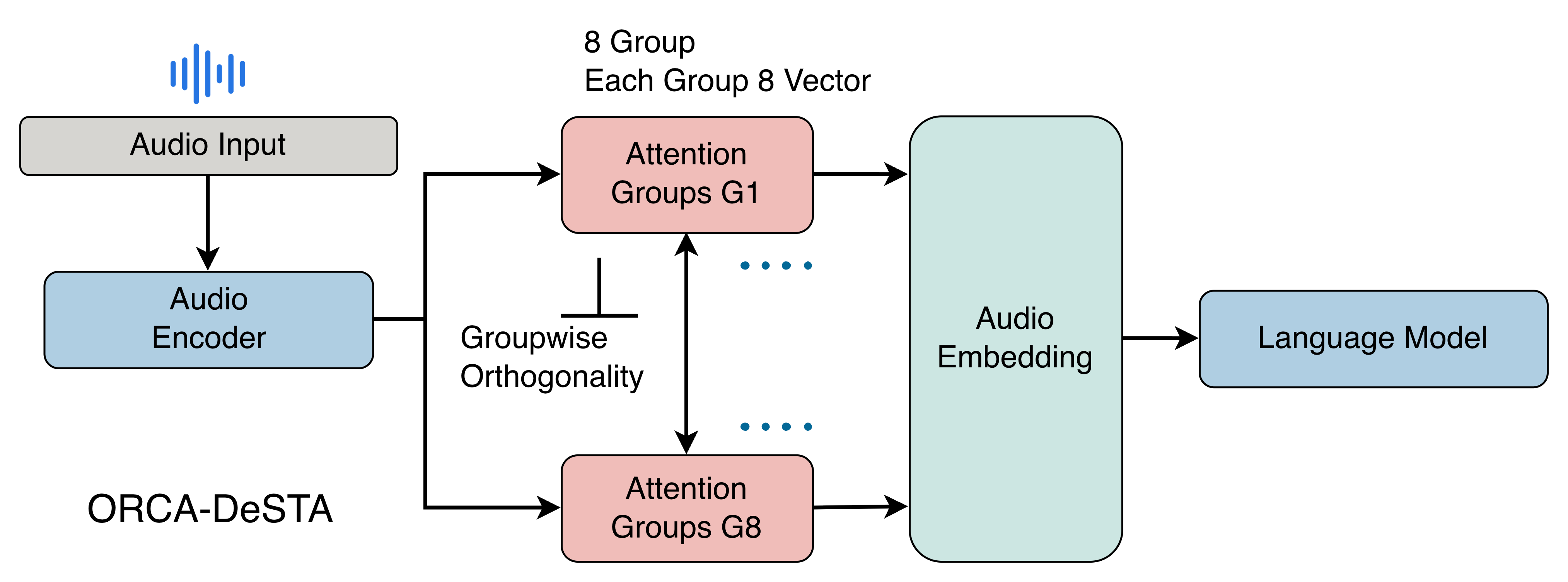}
\caption{\textbf{\orca{} at a glance.} \orca{} partitions the $K{=}64$ connector queries into $G{=}8$ groups of $J{=}8$ tokens. Each group has its own learnable queries and cross-attends to the encoder output through a shared Q-Former backbone. A regularizer pushes the group output centers toward mutually orthogonal directions, so the audio prefix for the LLM spans multiple subspaces instead of collapsing onto one.}
\label{fig:overview}
\end{figure}

Two properties of this design drive the collapse.
First, the loss in language-modeling rewards only the prediction of text, creating a convergence pressure that drives all $K$ query outputs toward a single direction.
Second, queries typically attend to a single encoder layer, and even multi-layer designs such as \desta{}~\cite{lu2025desta} share their layer weights across all queries, so no query can specialize to a different encoder depth.
\orca{} addresses both by partitioning the queries into $G$ groups whose outputs are regularized to be orthogonal and giving each group its own layer mixture (Fig.~\ref{fig:overview}).

The group structure is simple.
We divide $K$ queries into $G$ groups of $J$ tokens, $K=GJ$.
Each group has its own learnable queries $\mathbf{Q}_g$ but shares the Q-Former backbone weights, so the parameter count matches the baseline.
Each group also learns layer-attention weights $\alpha_{g,l}$ over a small set of encoder layers $\mathcal{E}=\{7,15,23,31\}$ (0-indexed), the same set used by \desta{}~\cite{lu2025desta}.
The output of group $g$ is
\begin{equation}
    \mathbf{Z}_g
    =
    \!\sum_{l\in\mathcal{E}}\!
    \alpha_{g,l}\,
    \mathrm{QFormer}(\mathbf{Q}_g,\mathbf{H}^{(l)}),
    \label{eq:group_output}
\end{equation}
and the complete output is $\mathbf{Z}=[\mathbf{Z}_1;\ldots;\mathbf{Z}_G]$.

A regularizer then shapes these groups, enforcing two properties of the connector's output.
First, different groups should point in different directions in the output space, so groups occupy non-overlapping subspaces, which targets query collapse directly.
Second, tokens within the same group should stay related but not identical, so each group can encode multiple facets of the same attribute without collapsing to a single point.

Let $\mathbf{z}_{g,j} \in \mathbb{R}^d$ be the $j$-th output token of group $g$ (the $j$-th row of $\mathbf{Z}_g$).
Let $\bar{\mathbf{z}}_g = \frac{1}{J}\sum_{j=1}^{J} \mathbf{z}_{g,j} \in \mathbb{R}^d$ be the group center.
We add two terms to the language-modeling loss:
\begin{equation}
    \mathcal{L}_{\mathrm{group}}
    =
    \lambda_{\mathrm{inter}}\!\!\sum_{g<g'}\!
    \!\!\left(\!\frac{\bar{\mathbf{z}}_g^\top \bar{\mathbf{z}}_{g'}}{\|\bar{\mathbf{z}}_g\|\,\|\bar{\mathbf{z}}_{g'}\|}\!\right)^{\!2}\!
    +
    \lambda_{\mathrm{intra}}\,\Omega_{\mathrm{intra}}.
    \label{eq:group_loss}
\end{equation}
The first sum is the \textit{inter-group term}.
It measures the squared cosine alignment between every pair of distinct group centers $(\bar{\mathbf{z}}_g, \bar{\mathbf{z}}_{g'})$.
When the centers are orthogonal the term is zero; when they collapse onto each other it grows toward one.
We square the cosine so its sign is irrelevant and the gradient stays smooth at zero, and use cosine rather than inner product so groups cannot escape the penalty by shrinking themselves.
The scalar weight $\lambda_{\mathrm{inter}} > 0$ controls how strongly this competes with the language-modeling loss.
This constraint is purely geometric: it rewards group centers for pointing in different directions and never references speaker, emotion, or any other attribute. Its role is to make the single-axis collapse geometrically infeasible, so that non-semantic variation has room to survive; it does not tell the model which attributes to encode, as the language-modeling loss remains the only task signal.

The second is the \textit{intra-group term} $\Omega_{\mathrm{intra}}$.
Without it, the inter-group penalty alone could push every group to a single point and waste the multi-token capacity inside each group.
We instead pick a target similarity $s^\star \in (0, 1)$ and penalize squared deviation of the average within-group similarity from $s^\star$:
\begin{equation}
    \Omega_{\mathrm{intra}}
    \;=\;
    \frac{1}{G}\sum_{g=1}^{G}
    \!\left(\!
    \frac{2}{J(J{-}1)}\!\sum_{j<j'}\!
    \frac{\mathbf{z}_{g,j}^{\!\top}\mathbf{z}_{g,j'}}{\|\mathbf{z}_{g,j}\|\,\|\mathbf{z}_{g,j'}\|}
    -s^\star\!
    \right)^{\!2}.
    \label{eq:intra}
\end{equation}
The inner double sum is the average pairwise cosine similarity among the $J$ tokens of group $g$.
The outer squared deviation makes the penalty symmetric: tokens that drift apart (similarity $<s^\star$) and tokens that collapse together (similarity $\to 1$) are both penalized.
The scalar weight $\lambda_{\mathrm{intra}} > 0$ controls the strength. Operating on group centers rather than every cross-group token pair keeps the constraint coarse and the gradient stable: tokens inside one group can still encode related aspects of the same attribute without being individually pushed away from tokens in other groups.

We instantiate \orca{} with $G{=}8$, $J{=}8$, and $K{=}64$, matching the baseline's token budget exactly, and set $\lambda_{\mathrm{inter}}{=}0.1$ and $\lambda_{\mathrm{intra}}{=}0.03$.
The target similarity $s^\star{=}0.3$ is moderately small but non-trivial: it keeps groups internally coherent while leaving room for token-level diversity inside each group.
The Q-Former has 6 layers, and the audio encoder and the LLM are frozen, so only the connector is trained.
We follow the public \desta{} recipe~\cite{lu2025desta}: 5 epochs ($\sim$250k steps) on AQA5M, Adam optimizer, peak learning rate $1{\times}10^{-4}$, cosine annealing with 2000-step warmup, global batch size 96, on 8$\times$A100-80GB GPUs.
The total loss is $\mathcal{L}=\mathcal{L}_{\mathrm{LM}}+\mathcal{L}_{\mathrm{group}}$.

The added cost is small.
The Q-Former backbone weights are shared across all groups, so relative to the single-group baseline the query-embedding budget is identical ($GJ{=}K{=}64$) and the two connectors match in parameter count up to a handful of layer-attention scalars; over a connector-free pipeline the added queries are only $GJd{\approx}0.065$M for $d{=}1024$.
The regularizer adds $G(G{-}1)/2$ cosine terms on mean vectors, and inference latency matches a standard Q-Former.

\section{Experiments}
\label{sec:experiments}

We answer three questions in order.
\textbf{(Q1)} Does fixing query collapse at the connector also reduce information loss, as the diagnosis predicts?
\textbf{(Q2)} Does this propagate to better paralinguistic reasoning on real tasks?
\textbf{(Q3)} Does the improvement survive acoustic noise?

We compare two 4B models that share the Whisper-Large-V3 encoder, the Qwen3-4B~\cite{yang2025qwen3} backbone, the AQA5M training data~\cite{lu2025desta}, and the training recipe.
The only difference is the connector: a standard Q-Former (the \desta{} 4B baseline) versus \orca{}.
For context, we also report scaling baselines: Qwen2.5-Omni~\cite{xu2025qwen2}, \qwenaudio{}~\cite{chu2024qwen2}, Audio Flamingo-3~\cite{goel2025audio}, and \desta{} (8B)~\cite{lu2025desta}.

AQA5M comprises about 7{,}000 hours of audio across 50 publicly available datasets, with broad coverage of paralinguistic metadata that lets the model learn to attend to non-semantic attributes during training (Table~\ref{tab:dataset_composition}).
Speech accounts for the majority ($\sim$77\%), with annotations spanning emotion, gender, age, accent, and prosodic features, while environmental sound ($\sim$14\%) and music ($\sim$7\%) round out the corpus.
Following the self-generation principle of \desta{}~\cite{lu2025desta}, training targets are generated by the same Qwen3-4B that serves as the backbone, which keeps the supervision and the model's native output style on the same distribution.

\begin{table}[t]
\centering
\caption{\textbf{AQA5M domain composition.} The dataset covers 50 source datasets across three audio domains. We list the metadata types available in each domain.}
\label{tab:dataset_composition}
\begin{small}
\resizebox{\linewidth}{!}{
\begin{tabular}{lccp{4.5cm}}
\toprule
\textbf{Domain} & \textbf{Hours} & \textbf{\#Datasets} & \textbf{Metadata Types} \\
\midrule
Speech & $\sim$5{,}400 & 38 & emotion, gender, age, accent, pitch, speaking rate, SNR, language, speaker ID \\
Env.\ Sound & $\sim$1{,}000 & 7 & event category, caption, duration \\
Music & $\sim$500 & 5 & genre, instrument, MIDI note, song title \\
\midrule
\textbf{Total} & $\sim$7{,}000 & 50 & \\
\bottomrule
\end{tabular}
}
\end{small}
\end{table}

We evaluate on two benchmarks.
SAKURA~\cite{yang2025sakura} measures multi-hop paralinguistic reasoning over gender, animal, language, and emotion attributes, and MMAU~\cite{sakshi2024mmau} measures general audio understanding over sound, speech, and music.
We use the official prompts and greedy decoding, and report accuracy.
The diagnostic experiments use CREMA-D~\cite{cao2014crema} with speaker-disjoint folds.

\subsection{Q1: Collapse and Variance Recovery}
\label{subsec:diagnostic_validation}

\begin{table}[t]
\centering
\caption{\textbf{Query redundancy by condition on CREMA-D.} The baseline collapses uniformly across emotions and sentences; \orca{} reduces the similarity uniformly as well. Values are reported as mean $\pm$ standard deviation across conditions.}
\label{tab:redundancy_breakdown}
\setlength{\tabcolsep}{4pt}
\begin{small}
\begin{tabular}{lcc}
\toprule
\textbf{Condition} & \textbf{\desta{} 8B} & \textbf{\orca{} 4B} \\
\midrule
Overall & 0.923 & \textbf{0.077} \\
Per emotion (6 classes) & $0.923 \pm 0.001$ & $0.077 \pm 0.001$ \\
Per sentence (12 sents.) & $0.923 \pm 0.001$ & $0.077 \pm 0.003$ \\
\bottomrule
\end{tabular}
\end{small}
\end{table}

\begin{table}[t]
\centering
\caption{\textbf{Re-running the diagnostics on \orca{}.} Only redundancy is in the loss; the variance recovery is a downstream effect.}
\label{tab:diagnostic_validation}
\setlength{\tabcolsep}{4pt}
\begin{small}
\resizebox{\linewidth}{!}{
\begin{tabular}{lcc}
\toprule
\textbf{Diagnostic} & \textbf{\desta{} 8B} & \textbf{\orca{} 4B} \\
\midrule
Query cosine sim.\ \textit{(targeted)} & 0.923 & \textbf{0.077} \\
Cross-speaker variance & 0.0015 & \textbf{0.113} \\
Discriminative margin $\Delta S$ & 0.010 & \textbf{0.085} \\
\bottomrule
\end{tabular}
}
\end{small}
\end{table}

We re-run the two diagnostics from Section~\ref{sec:diagnosis} on \orca{}'s connector outputs.
If query collapse drives information loss, fixing collapse should reduce information loss as well, even though no loss term addresses it directly.
Query cosine similarity drops from 0.923 to 0.077 (Table~\ref{tab:diagnostic_validation}), a 12$\times$ reduction.
The regularizer directly penalizes this quantity, so the reduction is expected.
The effect is uniform across emotions and sentences (Table~\ref{tab:redundancy_breakdown}), matching the uniformity of the original collapse.

The second diagnostic also improves: $\Delta S$ grows from 0.010 to 0.085 (Table~\ref{tab:diagnostic_validation}).
We also measure cross-speaker variance, the per-dimension variance of the mean-pooled connector output across speakers reading the same sentence, averaged over dimensions and sentences.
It rises from 0.0015 to 0.113, a 75$\times$ increase.
No loss term penalizes either quantity, so both improvements are downstream effects of reducing collapse alone.

\subsection{Q2: Paralinguistic Reasoning Downstream}
\label{subsec:downstream}

\begin{table*}[t]
\centering
\caption{\textbf{Main results (accuracy)} on SAKURA (paralinguistic reasoning) and MMAU (general audio understanding). \textbf{S}: single-hop, \textbf{M}: multi-hop. \textbf{Bold}: best in column; \underline{underline}: second best. The \desta{} 4B baseline and \orca{} share encoder, LLM, data, and training recipe; only the connector differs. With 4B parameters and 7k hours of audio, \orca{} ranks first on overall multi-hop reasoning (SAKURA Avg-M) and is first or second on 10 of the 14 reported metrics, whereas the Qwen and Audio~Flamingo models, trained on one to three orders of magnitude more audio, lead on single-hop but drop sharply under multi-hop.}
\label{tab:downstream}
\resizebox{\textwidth}{!}{%
\begin{tabular}{lcc cccccccccc cccc}
\toprule
& & & \multicolumn{10}{c}{\textbf{SAKURA (Reasoning)}} & \multicolumn{4}{c}{\textbf{MMAU (General)}} \\
\cmidrule(lr){4-13} \cmidrule(lr){14-17}
& & & \multicolumn{2}{c}{\textbf{Animal}} & \multicolumn{2}{c}{\textbf{Gender}} & \multicolumn{2}{c}{\textbf{Emotion}} & \multicolumn{2}{c}{\textbf{Lang}} & \multicolumn{2}{c}{\textbf{Avg}} & & & & \\
\textbf{Model} & \textbf{Param} & \textbf{Audio (hrs)} & \textbf{S} & \textbf{M} & \textbf{S} & \textbf{M} & \textbf{S} & \textbf{M} & \textbf{S} & \textbf{M} & \textbf{S} & \textbf{M} & \textbf{Sound} & \textbf{Speech} & \textbf{Music} & \textbf{Avg} \\
\midrule
\multicolumn{17}{l}{\textit{Reference systems (varied scale \& data)}}\\
Qwen2.5-Omni & 3B & $\sim$1.7M--3.3M est. & \textbf{92.2} & \underline{70.6} & 64.8 & 22.6 & 27.4 & 15.4 & 80.0 & 29.0 & 66.1 & 34.4 & \underline{70.0} & 56.5 & \underline{59.3} & 61.9 \\
\qwenaudio{}  & 7B & $\sim$520k & \underline{90.0} & 60.6 & \underline{89.0} & 43.4 & 62.2 & 40.2 & 83.4 & 46.4 & \textbf{81.2} & 47.7 & 54.4 & 43.8 & 55.4 & 51.2 \\
Audio Flamingo-3   & 8B & $\sim$100k & 88.0 & 61.6 & 66.4 & 49.4 & \textbf{70.8} & 41.2 & 93.8 & 43.6 & 79.8 & 49.0 & \textbf{80.2} & 67.6 & \textbf{71.9} & \textbf{73.2} \\
\desta{} (8B) & 8B & 7{,}000 & 58.8 & 58.0 & \textbf{89.2} & \underline{84.0} & 62.2 & \textbf{57.4} & \textbf{96.4} & \underline{80.0} & 76.7 & \underline{69.9} & 64.6 & 64.9 & 54.8 & 61.4 \\
\midrule
\multicolumn{17}{l}{\textit{Controlled comparison: same encoder, LLM, data \& recipe (4B, 7k-hr); only the connector differs}}\\
\desta{} 4B Baseline     & 4B & 7{,}000 & 30.8 & 34.2 & 44.2 & 38.0 & 56.0 & 50.8 & 93.0 & 72.2 & 56.0 & 48.8 & 58.6 & \textbf{74.2} & 47.9 & 60.2 \\
\rowcolor{gray!10} \textbf{\orca{} (ours)} & 4B & 7{,}000 & 76.0 & \textbf{76.2} & 87.6 & \textbf{85.8} & \underline{63.8} & \underline{57.2} & \underline{94.4} & \textbf{81.6} & \underline{80.5} & \textbf{75.2} & 64.0 & \underline{71.8} & 52.4 & \underline{62.7} \\
\bottomrule
\end{tabular}%
}
\end{table*}

Table~\ref{tab:downstream} gives the main numbers.
Existing models struggle on paralinguistic reasoning even when recognition is strong.
Qwen2.5-Omni scores 22.6\% on Gender multi-hop and 15.4\% on Emotion multi-hop despite high automatic speech recognition (ASR) accuracy, so the failure is not in recognition but in preserving non-semantic information through alignment.

The connector alone explains a large part of this gap.
With identical encoder, LLM, data, and recipe, \orca{} (4B) beats the \desta{} 4B baseline on every multi-hop category by large margins: Animal 76.2 vs.\ 34.2, Gender 85.8 vs.\ 38.0, Emotion 57.2 vs.\ 50.8, and Language 81.6 vs.\ 72.2, for a +26.4-point gain in multi-hop average.
Since the only differences are the orthogonal group partition and the per-group layer mixtures, this isolates the effect to the connector.
\orca{} (4B) also exceeds the 8B \desta{} by +5.3 points on multi-hop average, though this comparison is not controlled, as the two models use different LLM backbones.

The gains are not uniform, and the ordering tracks how much each attribute relies on non-semantic acoustic structure.
They are largest for Gender (+47.8) and Animal (+42.0), both carried by spectrally broad cues that a collapsed connector destroys, smaller for Language (+9.4), whose phonemic identity partially survives collapse, and smallest for Emotion (+6.4), which depends on fine prosodic cues that are hard to preserve through any compression.

The same ordering holds, more weakly, on the single-hop split.
\orca{} wins there too (80.5 vs.\ 56.0 average against the 4B baseline; 80.5 vs.\ 76.7 against the 8B), but by a smaller margin.
Single-hop questions ask the model to read an attribute directly, which even a partially collapsed connector can sometimes do through surface cues; multi-hop questions require combining it with internal knowledge, exactly where carrying the attribute through alignment matters.

Figure~\ref{fig:case_study} shows what information loss looks like at the output level.
Given a happy CREMA-D utterance, the \desta{} 8B baseline misclassifies the emotion as ``Fear'' and then builds a coherent story around that wrong choice, complete with invented acoustic details.
With cross-speaker variance crushed by the collapsed connector, the LLM has no acoustic content to ground its description in, so it fills in plausible-sounding details from the wrong emotion choice, a failure mode documented for audio-language models that hallucinate ungrounded reasoning~\cite{aha2025}.
The same prompt with \orca{} produces a description consistent with the actual delivery.

\begin{figure}[t]
\centering
\fbox{\parbox{0.94\columnwidth}{\footnotesize
\textbf{Audio:} CREMA-D 1001\_TAI\_HAP\_XX. Male speaker says ``The airplane is almost full'' in a \textbf{happy}, upbeat tone.\\[2pt]
\textbf{Prompt:} ``What is the emotion of the speaker? Describe the audio.''\\[3pt]
\textbf{\desta{} 8B:} \textit{``The emotion expressed in the audio is \textbf{Fear}. The speaker's voice conveys a sense of anxiety or apprehension, suggesting unease about the situation. The vocal delivery is likely high-pitched and rapid, with a slight tremble, indicating heightened emotional arousal. Given the line `The airplane is almost full,' the speaker may be expressing nervousness about boarding a crowded flight.''}\\[3pt]
\textbf{\orca{} (4B):} \textit{``The emotion expressed by the speaker is \textbf{happy}. The audio features a male voice (age 22) delivering `The airplane is almost full' with a positive and upbeat emotional inflection, suggesting enthusiasm or excitement. The voice quality is clear and energetic, consistent with the cheerful delivery.''}
}}
\caption{\textbf{Downstream effect of connector collapse on emotion description.}
Given the same audio and prompt, \desta{} 8B misclassifies the emotion and invents acoustic details (tremble, high pitch) to match, while \orca{} correctly identifies the emotion and describes the actual delivery.
Attribute details such as the stated age remain unverified model guesses; the example illustrates emotion grounding rather than attribute-level fidelity.}
\label{fig:case_study}
\end{figure}

On general audio understanding the picture is different.
On MMAU, \orca{} (62.7\%) is on par with Qwen2.5-Omni (61.9\%) and the 8B baseline (61.4\%), while Audio Flamingo-3 (73.2\%) leads with a different paradigm and more data.
Structure helps paralinguistic reasoning; scale helps broad audio knowledge; the two are complementary.
\orca{} keeps Speech accuracy high (71.8\%, second only to the 4B baseline; the small regression suggests the orthogonality constraint shifts some capacity from general speech understanding toward paralinguistic subspaces) and matches the field on Sound and Music, where success depends more on the breadth of audio events seen during pretraining~\cite{gemmeke2017audioset,kim2019audiocaps,piczak2015esc,engel2017nsynth} than on whether the connector preserved paralinguistic structure.

\subsection{Q3: Robustness Under Noise}
\label{subsec:robustness}

\begin{table}[t]
\centering
\caption{\textbf{Robustness.} SAKURA multi-hop accuracy (\%) under additive Gaussian noise. $\Delta_{10}$ is the change from clean to a 10\,dB signal-to-noise ratio (SNR).}
\label{tab:robustness}
\setlength{\tabcolsep}{3pt}
\begin{small}
\begin{tabular}{lccc|ccc}
\toprule
& \multicolumn{3}{c|}{\textbf{\desta{} 8B}} & \multicolumn{3}{c}{\textbf{\orca{} 4B}} \\
\textbf{Category} & Clean & 10\,dB & $\Delta_{10}$ & Clean & 10\,dB & $\Delta_{10}$ \\
\midrule
Animal   & 58.0 & 55.6 & {\color{gray}$-$2.4}  & 76.2 & 59.8 & {\color{gray}$-$16.4} \\
Gender   & 84.0 & 68.6 & {\color{gray}$-$15.4} & 85.8 & 85.0 & {\color{gray}$-$0.8} \\
Emotion  & 57.4 & 39.0 & {\color{gray}$-$18.4} & 57.2 & 41.0 & {\color{gray}$-$16.2} \\
Language & 80.0 & 78.6 & {\color{gray}$-$1.4}  & 81.6 & 83.8 & {\color{gray}$+$2.2} \\
\midrule
\textbf{Avg} & 69.9 & 60.5 & {\color{gray}$-$9.4} & \textbf{75.2} & \textbf{67.4} & {\color{gray}$-$7.8} \\
\bottomrule
\end{tabular}
\end{small}
\end{table}

We add white Gaussian noise at 10\,dB SNR and re-evaluate (Table~\ref{tab:robustness}).
Gender shows the clearest separation.
Both models start near 84--86\% on clean audio, yet \desta{} 8B drops 15.4 points multi-hop while \orca{} drops 0.8.
Gender cues such as fundamental frequency (F0) and formant structure are spectrally broad and survive noise, so the gap reflects whether the connector preserves them.
The diagnostics in \S\ref{sec:diagnosis} show the baseline does not preserve them, consistent with its sharp drop under noise, whereas \orca{} retains a usable acoustic pathway.
On Language, both models are nearly invariant, since phonemic and rhythmic cues are robust by nature.
On Animal, both drop and \orca{} drops further from a higher start, the generic regression any model shows once the absolute signal is degraded.
Emotion is intrinsically fragile for both models because fine prosodic cues are masked by additive noise at the signal level.
Even at 10\,dB, \orca{} retains 67.4\% multi-hop average, just 2.5 points below the \desta{} 8B \textit{clean} average.
Structure cannot rescue physically corrupted cues, but it makes a real difference for cues the connector would otherwise discard.

\subsection{Group Specialization}
\label{subsec:layer_attention}

\begin{figure}[t]
\centering
\includegraphics[width=0.95\linewidth]{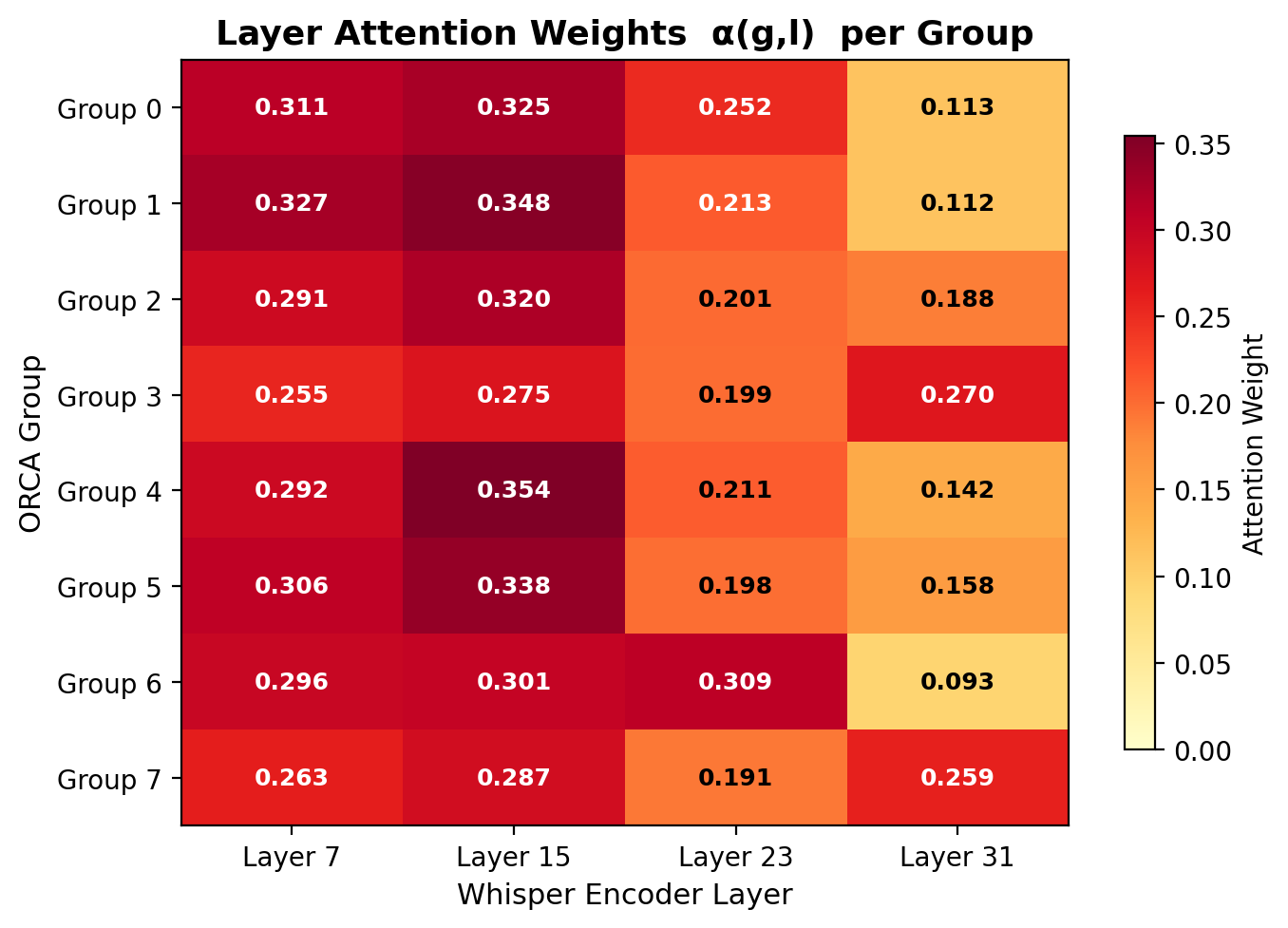}
\caption{\textbf{Learned layer-attention weights} $\alpha_{g,l}$. Rows are the 8 groups, columns are Whisper layers $\{7,15,23,31\}$ (0-indexed). Without supervision, groups split into shallow-, mid-, and deep-layer specialists.}
\label{fig:layer_attn}
\end{figure}

We never tell the groups what to learn.
Yet the learned $\alpha_{g,l}$ split the eight groups into three patterns (Fig.~\ref{fig:layer_attn}).
Groups 0, 1, 4, and 5 act as shallow specialists, concentrating on layers 7 and 15, which encode lower-level spectral and prosodic features; Group 4, for instance, places $\alpha{=}0.354$ on layer 15 and $0.292$ on layer 7, with low weight on the deep layers.
Group 6 is a mid-layer specialist, uniquely peaking on layer 23 ($\alpha{=}0.309$) and assigning the lowest weight to the final layer 31 ($0.093$) among all groups.
Groups 3 and 7 are deep specialists, placing the most weight on the final layer 31 of any group ($\alpha{=}0.270$ and $0.259$), about $2\times$ that of the shallow specialists.
Read against layer-wise analyses of speech encoders~\cite{hsu2026anatomy,pasad2021layer,baevski2020wav2vec,hsu2021hubert}, this specialization suggests that orthogonality on output centers pushes groups to draw from different processing depths, a natural way to occupy non-overlapping output subspaces.
The structural inductive bias does the work; specialization emerges for free, echoing how group structure can yield specialized parts without part-level supervision in other domains~\cite{xu2022groupvit}.

\subsection{Discussion}
\label{subsec:limits}

The fix is structural, so its reach is bounded by what is upstream of the connector.

One bound is the encoder, which sets a floor.
\orca{} cannot recover information the audio encoder has already discarded, nor can it extract cues that are physically masked by noise.
Emotion is the clearest example: the clean-condition gain over the 4B baseline is the smallest of the four SAKURA categories (+6.4 vs.\ +47.8 for Gender), and the 10\,dB drop is similar for both \orca{} and the 8B baseline.
Fine prosodic cues such as pitch contour, voice quality, and timing live close to the noise floor of any frame-level feature extractor, and structure cannot push them above it.

A second bound is the loss.
Group orthogonality opens room for non-semantic axes, but it does not change which axis the language-modeling loss most heavily rewards.
The semantic axis remains primary by construction, and other groups specialize only to the extent that the gradient finds them useful.
This bounds what an architecture-only fix can achieve and points to data and task design as complementary levers.

\section{Conclusion}

Q-Former connectors in standard audio-language models collapse their
output onto a single semantic direction: the mean pairwise cosine
similarity among $K{=}64$ output vectors reaches 0.923 on the \desta{}
baseline, and the cross-speaker discriminative margin is only
$\Delta S{=}0.010$.
ORCA addresses this upstream cause by partitioning the $K$ queries into
$G$ orthogonal groups, each with its own learned encoder-layer mixture,
at no additional parameter or inference cost.
The geometric constraint reduces query cosine similarity from 0.923 to
0.077 ($12{\times}$) and raises cross-speaker variance by $75{\times}$,
both as downstream effects without any attribute supervision.
On SAKURA multi-hop reasoning, ORCA gains 26.4 points over the
identically trained 4B baseline (75.2\% vs.\ 48.8\%), and group
specialization in encoder depth emerges without supervision.

The result illustrates a general principle.
When a training objective rewards only one mode of variation, the
connector discards everything else.
Adding per-attribute losses treats symptoms one at a time and requires
annotated targets for each attribute.
Reshaping the output geometry so that multiple subspaces coexist in
the bottleneck addresses the structural cause, and gains propagate to
attributes the designer did not specify.
Connector geometry is therefore a first-class design axis alongside
model scale and training data: a 4B model with the right structure can
recover what an 8B model with a collapsed connector discards.

\bibliographystyle{IEEEtran}
\bibliography{references}

\end{document}